\documentclass{aastex}
\usepackage{spr-astr-addons}
\usepackage{verbatim}

\RequirePackage{color}

\newcommand{\sect}[1]{\setcounter{equation}{0}\section{#1}}
\renewcommand{\theequation}{\arabic{section}.\arabic{equation}}
\newcommand{\bfm}[1]{\mbox{\boldmath${#1}$}}

\begin{document}

\title{Analytical predictions of non-Gaussian distribution\\ parameters for stellar plasmas}
\shorttitle{Analytical predictions of non-Gaussian distribution parameters}
\shortauthors{A.M. Scarfone {\em et al}.}

\author{A.M. Scarfone}
\affil{Ist.
Naz. di Fisica della Materia (C.N.R. -- I.N.F.M.) and Dipart. di Fisica,\\ Unit\'a del Politecnico di Torino, I-10129 Torino, Italy.}
\and
\author{P. Quarati}
\affil{Dipart. di Fisica del Politecnico di Torino,
I-10129 Torino, Italy and\\ Ist. Naz. Fisica Nucleare (I.N.F.N.) Cagliari,
I-09042 Monserrato, Italy}
\and
\author{G. Mezzorani}
\affil{Dipart. di Fisica dell'Universit\`a di Cagliari,
I-09042 Monserrato, Italy and\\ Ist. Naz. Fisica Nucleare (I.N.F.N.) Cagliari,
I-09042 Monserrato, Italy}
\and
\author{M. Lissia}
\affil{Ist. Naz. Fisica Nucleare (I.N.F.N.) Cagliari,
I-09042 Monserrato, Italy}

\date{\today}

\begin {abstract}
Stimulated by the recent debate on the physical relevance and
on the predictivity of $q$-Gaussian formalism, we present
specific analytical expressions for the parameters characterizing
non-Gaussian distributions, such as the nonextensive parameter $q$,
expressions that we have proposed for different physical systems, an
important example being plasmas in the stellar cores.\\

\end {abstract}
\maketitle

\sect{Introduction} Many recent experimental measurements of space
and momentum distribution functions of particle systems and of
physical quantities in several fields of natural and social science
show behaviors that can be described with non-Gaussian
distributions: from cosmic rays to dark matter, from astrophysical
systems to turbulence and quark-gluon plasmas, from biology to
finance. More concisely: power laws are detected in several complex
systems (Gell-Mann and Tsallis 2004).

Among various non-Gaussian distributions that
emerge from consistent thermodynamical and statistical frameworks
 (Kaniadakis et al. 2004; Kaniadakis et al. 2005; Lavagno et al. 2007),
$q$-Gaussians, based on the so-called nonextensive statistical
mechanics introduced by Tsallis (1988), are appealing for their
simplicity and have found the most applications.
A critical debate on its theoretical foundation and on its
practical applicability is still going on. In particular, there
has been a great effort in this field to
construct a statistical mechanics capable of describing systems
affected by nonlocal and memory effects. Relevance of these problems in astrophysics
and space science has been widely discussed in many papers
Saxena et al. (2004,2006) (see also Mathai and Haubold 2007; Haubold and Kumar 2007
for comments on applications of $q$-Gaussians in the above field).

Two important questions regarding
$q$-Gaussian distributions have been recently raised by Douxois (2007):
if $q$-Gaussian law describes physical
phenomena and if the entropic $q$ parameter could be predicted in
terms of microscopic parameters. Tsallis (2008), has
answered these questions reviewing the present state of nonextensive
statistical mechanics and reporting analytical relations for the
parameter $q$ that have appeared over the recent years. It is our opinion that natural
field of application and validation of $q$-Gaussian distributions is astrophysics and space science.

In this work, we present additional interpretations of the non-Gaussian
parameters, mostly in terms of the parameter $q$,
with the corresponding analytical expressions, that our group have
proposed: a general statistical
interpretation based on Fokker-Plank kinetic equations and on Langevin equation,
phenomenological interpretations about weakly nonideal plasmas
and charged particles in electric and magnetic fields inelastically
interacting with a medium and an interpretation based on the
quasi-particle description of correlated particles with finite lifetime.
This list is not exhaustive, but we believe it offers an
important contribution to the scientific debate on the
physical interpretation of the deformation and
on the possibility of directly linking parameter $q$, or other
analogous parameters, to
the underlying microphysics; in these specific cases, the
link has simple analytical forms.
In the cases presented, the deviation of the distributions from
the exponential case, $q\neq 1$, is due to the non-trivial spatial
correlations among the particles of the system or time correlations
of long-living metastable systems. The possibility of a two-parameter
distribution that capture the power-law behavior in two different
asymptotic regimes (Kaniadakis et al. 2004, 2005)
and its application to chaotic systems is also recalled (Tonelli et al. 2006).

The physically important case of plasmas in stellar cores is
discussed to clarify the importance of non-Gaussian distributions.


\sect{Non-Gaussian distribution parameters}

By solving a nonlinear Fokker-Planck equation with dynamical and
friction coefficients given by a polynomial expression with positive
coefficients, in 1992 we derived (Kaniadakis and Quarati 1993)
a set of stationary non-Maxwellian distributions with
depleted and enhanced tails compared to Maxwellian tail.

Analytical forms of the distributions were given for expansion
up to the sixth order, with  the parameters of the distributions
expressed in terms of the derivatives of the dynamical friction
and diffusion coefficients. All these distributions deviates from
the Maxwellian one, which is recovered in the limit of
vanishing nonlinear corrections to the Fokker-Planck equation.

As a specific example, Eq.~(30b) of Kaniadakis and Quarati (1993)
\begin{equation}\label{eq:30b}
    f(u) = \frac{\Gamma(\mu)}{\sqrt{\pi} \Gamma(\mu-1/2)}
      \left( 1+\frac{u^2}{\mu-1}\right)^{-\mu}
\end{equation}
with $\mu=1+\beta_0/(2\gamma_1)$, where $\beta_0$ and $\gamma_1$
are the coefficients of the leading and subleading term of the
expansion in the velocity of the dynamical friction and diffusion
terms.

Later on, we realized that this specific distribution, which in
our paper we called second-order stationary solution of the
Fokker-Plank equation, basically coincides with the
$q$-distribution obtained independently by Tsallis a few
years earlier with an entropic approach (Tsallis 1988).
In this case the analytical form of corresponding
entropic parameter $q$ is
\begin{equation}\label{eq:qFK}
    q= 1+\frac{1}{\mu} = 1+ \frac{2\gamma_1}{2\gamma_1+\beta_0}\quad .
\end{equation}

Clearly, non-Gaussian distributions and extended entropies were
already known and used in different fields (Mathai and Rathie 1975; Haubold et al. 2007; Kaniadakis and Lissia 2004), however it is only after Tsallis seminal paper
that systematic extensions of thermodynamics and statistical
mechanics were developed. Such approaches
received  great attention and are actually used in many fields in science.

In Kaniadakis and Quarati (1997), we have investigated more in detail the
relationship between our results and the nonextensive statistical
mechanics results, realizing the connection between the physical
meaning of some of our results of Kaniadakis and Quarati (1993) and the corresponding
Tsallis distribution and entropic parameter $q$.

We have also studied Kaniadakis et al. 2004, 2005) a wider class of generalized two-parameter
statistical mechanics that includes not only the $q$-thermostatistics,
but also other physically motivated extensions such as the $\kappa$
distribution
with its appealing relativistic structure (Kaniadakis 2002, 2005),
and the Abe distribution related to quantum group (Abe 1997).
The interpretation of the two parameters exist only for selected cases or
when there remains only one effective parameter and it is still object
of study.


\subsection{Polynomial expansion of diffusion and drift coefficients}

One of the distributions emerging as equilibrium
solutions of the nonlinear Fokker-Plank equation for a specific form
of the coefficients was
first interpreted in term of $q$-distribution in
1997 (Kaniadakis and Quarati 1997),
where we gave an analytical expression of $q$ in terms of the
ratio of the noise induced drift over total drift for specific model with
leading order terms in a velocity expansion:
\begin{equation}
q={4\,\theta-1\over2\,\theta-1} \ ,
\end{equation}
where
\begin{equation}
\theta={J_{\rm n}(v)\over J_{\rm n}(v)+J_{\rm d}(v)} \ ,
\end{equation}
with $J_{\rm n}(v)={1\over2}\,\gamma\,(1-q)\,v$, $J_{\rm
d}(v)={1\over2}\,(3-q)\,v$ and $\gamma$ is a constant. The drift
coefficient $J(v)$ is the sum of the deterministic drift $J_{\rm
d}(v)$ plus the induced-noise drift $J_{\rm n}(v)$.\\
When the noise-induced drift is absent we have $q=1$  and the $q$
nonextensive distribution reduces to Maxwellian one. The
microscopic Langevin equation related to Tsallis distribution is
\begin{eqnarray}
\nonumber
& &{d\,v\over
d\,t}+\gamma\,\left[1-(1-q)\,\left(1+{\sigma\over2}\right)\right]\,v\\
\nonumber
&=&\left\{{\gamma\over
m\,\beta}\left[1-(q-1)\,\beta\,\gamma\right]+{\gamma\over2}\,
(q-1)\,v^2\right\}^{1/2}\!\!\Gamma(t)
\ ,\\
\end{eqnarray}
and the solution can easily be written in
\begin{equation}
v(t)=\overline v\left[{1\over\sqrt{\overline
t}}\int\Gamma(t)\,dt+{\rm arcsinh}\left({v(0)\over \overline v}\right)\right]
\ ,
\end{equation}
where $\Gamma(t)$ is a random Gaussian variable with zero mean value
and $\delta$ correlation function, $\overline v=2\,[1-(\overline
q-1)\,\beta\,\lambda]/[\beta\,m\,(\overline q-1)]$, $\overline
t=2/[\gamma\,(q-1)]$, $\overline q=(\sigma+4)/(\sigma+2)$ and
$\lambda$ is an energy dimensional normalization constant.

\subsection{Phenomenological derivation for correlated particles in plasmas}

Soon after the interpretation in terms of drift and diffusion
coefficients, it was possible
to derive an analytical expression for $q$ concerning stellar/solar
core plasmas, where particles are correlated and entangled to the
electromagnetic fields, and their interaction cannot be described
by local two-body potentials (Coraddu et al. 1999).
This first specific analytical expression of $q$ in terms of microscopic
parameters, derived from the solution of a Boltzmann equation valid
when deformation of the global thermodynamic equilibrium
distribution (Maxwellian) is small (about $0.8<q<1.2$), concerns the
ionic components of a weakly nonideal neutral plasma, as
the solar core, in presence of an electric microfield distribution.

The parameter $q$ can be given in terms of plasma parameter
$\Gamma =(Ze)^2 n^{1/3}$/(kT), the ratio of the mean (Coulomb)
potential energy and the mean kinetic (thermal) energy, where
$n$ is the average density,
and of a parameter $\alpha$ of the ion-ion correlation function
(whose value is $0.4<\alpha<0.9$), as defined by Ichimaru (Yan and Ichimaru 1986; Ichimaru 1992).
By fixing $q$, we can describe with a simple expression the
distribution of a system of correlated particles interacting
elastically not only through a pure Coulomb potential but also, for
instance, through a reinforced Coulomb interaction.

\subsubsection{The effects of random microfields in plasmas.\\}
 The deviation from Gaussian distribution in plasmas (Coraddu et al. 1999)
 has been recently generalized (Ferro and Quarati 2005) to consider the
presence of a microfield distribution of a random force $F$. The expression
of $q$ valid for small deformations, $0.8<q<1.2$ is
\begin{equation}
q=1\mp{F^2\over k\,\mu^2\,n^2}\,{\alpha_1^2\over\alpha_0^4} \ ,
\end{equation}
where the two signs indicates subdiffusion or superdiffusion, $k$ is
the energy transfer coefficient, $n$ is the number density, $\mu$ is
the reduced mass, $\alpha_0$ and $\alpha_1$ are dimensional
constants related to the particle interactions and correlations
through the cross sections $\sigma_0(v)=\alpha_0/v$ and
$\sigma_1(v)=\alpha_1$ and therefore to the collision frequencies
$\nu_0$ and $\nu_1(v)$.
If $F$ is proportional to the electric field $E$ ($F=e\,E$) and $E$
is greater than a given critical value $E_{\rm crit}$, then we obtain
\begin{equation}
q=1\pm2\,\delta \ ,\quad\quad{\rm valid \ for \ } \delta < 1/2\quad,
\end{equation}
where
\begin{equation}
\delta  = 12\,\Gamma^2\,\alpha^4 \ ,
\end{equation}
with $\alpha$ the above mentioned correlation parameter.

More concisely, we can derive that the correction $\delta$ is the square
of the ratio between the thermal energy density over the reinforced
energy density times the energy transfer coefficient.

For stellar cores, considering the different ionic components one
finds that very small corrections to the value $q=1$ of the order,
let us say, of a few percents can sensibly change the
thermonuclear fusion rates with important consequences on the
processes occurring in the core and leaving unchanged the bulk
properties of the star (Coraddu et al. 1999).
In the case of the sun, helioseismology
can test the proton distribution function (Degl'Innocenti et al. 1998).

\subsubsection{Electromagnetic fields in two-energy-level medium.\\}
 Rossani and Scarfone (2000) have self-consistently derived from a
linear Boltzmann equation an analytical expression of the $q$
parameter as an explicit function of the electric and magnetic field
when charged particles are inelastically interacting with a medium
endowed with two energy levels. Concisely
\begin{equation}
q=1+{4\over3}\,k\,T\,f({\bfm e},\,{\bfm b}) \ ,
\end{equation}
with
\begin{equation}
f({\bfm e},\,{\bfm b})={1\over(\Delta E)^2}\,{({\bfm e}\cdot{\bfm b})^2
+c_\ast\,{\bfm e}^2\over c_0\,c_\ast\,(c_\ast^2+{\bfm
b}^2)} \ ,
\end{equation}
where ${\bfm e}=q\,{\bfm E}/m$ and ${\bfm b}=q\,{\bfm B}/m$ are the
rescaled electric and magnetic fields, $c_0$ and $c_\ast$ are constant coefficients
related to the microscopic cross section (due to
the inelastic contributes), and $\Delta E$ is the
energy gap between the excited and the fundamental level of the
medium.

\subsubsection{Particle correlation function and phase space cell.\\}
The expression of the square fractional deviation of the number
density from a Maxwell-Boltzmann distribution of a system of
correlated particles by means of the evaluation of the phase space
volume occupied by a nonextensive system of $N$ classical particles
that slightly deviates from Maxwell-Boltzmann phase space volume
gives the possibility to derive an analytical expression of $q$ in
terms of the radial correlation function $g(r)$ (Quarati and Quarati 2003).

By considering the square fractional deviations from Maxwell-Boltzmann distributions defined by $\sum_i{(\delta n_i)^2\over n_{i_0}}$, where $\delta n_i$ is the variation of the number distribution and $n_{i_0}$ is the number distribution proportional to Boltzmann factor and using the relation (Goodstein 1975)
\begin{equation}
\sum_i{(\delta n_i)^2\over n_{i_0}}=1+{N\over
V}\int\left[g(r)-1\right]\,dr \ ,
\end{equation}
where the sum is extended over all energy levels, neglecting
large deviations, we have derived the expression:
\begin{equation}
q=1-{1\over9}\left\{{1\over N}+{1\over
V}\int\left[g(r)-1\right]\,dr\right\} \ .
\end{equation}
Therefore, when $g(r)\to1$ and $N\to\infty$ we have that $q\to1$.
Otherwise if $N\to\infty$ but the particles are correlated because
we take $g(r)\not=1$ we obtain
\begin{equation}
q=1-{1\over9\,V}\int\left[g(r)-1\right]\,dr \ ,
\end{equation}
and if the particles are not correlated [$g(r)=1$] but the number
$N$ is small and finite we have
\begin{equation}
q=1-{1\over9\,N} \ .
\end{equation}

\subsubsection{Fluctuations of collective effective parameter.\\}
 Correlations among particles and fluctuations of intensive
parameters, such as the inverse Debye-H\"{u}ckel radius,
$1/R_{\rm DH}$, produce nonlinear effects inducing the system in a
stationary metastable state that can be described by a
$q$-Gaussian distribution. It is possible to show
(Quarati and Scarfone 2007) that,
following the development usually adopted in superstatistics for the
inverse temperature(Beck 2001; Wilk and Wlodarczyk 2007), a neutral astrophysical plasma, as
the one of a stellar core, has a parameter $q$ defined by
\begin{equation}
q=1-\left[{\Delta(1/R_{\rm DH})\over1/R_{\rm DH}}\right]^2 \ ,
\end{equation}
where $\Delta(1/R_{\rm DH})$ is the fluctuation of the inverse
Debye-H\"{u}ckel radius. If we know the equation of state of the system
the correction factor can be  expressed in terms of temperature
fluctuation.

\subsection{Quasi-particle life-time and non-Gaussian momentum distribution}

Many properties of  a system of interacting particles can often be
described by weakly interacting excitations or quasi-particles.

Let us define
\begin{equation}
\Sigma(\omega,\,{\bfm p}^2)=\Sigma_{\rm R}+i\,\Sigma_{\rm I} \ ,
\end{equation}
the self energy of the one-particle propagator, where $\omega$ and
$\bfm p$ are related by $\omega={\bfm
p}^2/2\,m+\Sigma(\omega,\,{\bfm p}^2)$.\\
When the imaginary part of the self-energy cannot be disregarded
($\Sigma_{\rm I}>0$) the momentum distribution deviates from a Maxwellian
distribution, even if the energy itself is Gaussian.
By limiting ourselves to the case of $q>1$ for small $\Sigma_{\rm I}>0$
we have derived (Coraddu et al. 2000) a phenomenological
interpretation of the parameter $q$ as
\begin{equation}
q=1+C\,\left({\Sigma_{\rm I}\over\epsilon_{\rm p}}\right)^2 \ ,
\end{equation}
where $C$ is a constant, $\epsilon_{\rm p}={\bfm
p}^2/2\,m^\ast+\Sigma_{\rm R}$ and
$m^\ast=m\,[1-(\partial\,\Sigma_{\rm
R}/\partial\,\omega)]/[1+2\,m\,(\partial\,\Sigma_{\rm
R}/\partial\,{\bfm p}^2)]$.\\
Spatial and temporal correlations among ions have large effects on
thermonuclear reactions that occur between high-energy ions
tunneling the Coulomb barrier.
Quark-gluon plasma is also formed in heavy-ion collision and
similar plasma effects are important in the evaluation of reaction
rates. This non-Gaussian momentum distribution arising from the
finite lifetime of the quasi-particles  should always be taken
into account when studying nuclear processes in any plasma,
the more so when the tail of the momentum distribution is
important.


\sect{Two-parameter statistical mechanics: $\bfm\alpha$ and $\bfm\beta$.}

Entropies or power-law distributions of
many anomalous system entropies might be more completely described
by a two-parameter class of
logarithms (Kaniadakis et al. 2004, 2005; Tonelli et al. 2006).
In fact, the two parameters basically characterize the
possibly different power-law behaviors of the distribution for
large positive and negative arguments.

As an example, at the onset of chaos the logistic map shows an
universal power-law behavior that fixes one of the exponents. This
same exponent characterizes the time dependence of the entropy and
the sensitivity to initial conditions: one measures one of the two
and can predict the behavior of the other. The second exponent can
be related to the ratio of growth of entropies in different
formulations (Tonelli et al. 2006).

This generalized entropies and distributions can be introduced by
means of a corresponding  two-parameter class of logarithms
\begin{equation}\label{eq:logGen}
    \widetilde{\log}(\xi) \equiv \frac{\xi^{\alpha}-\xi^{-\beta}}{\alpha+\beta}
    \quad ,
\end{equation}
and an
infinity of one-parameter deformed logarithms can be obtained as
particular cases  by imposing relationships between the two parameters.

In some cases the
emerging deformed parameter can be related to the microscopic
quantities of the system under inspection. The $q$-statistical
mechanics  is recovered, as one-parameter subclass of
the two-parameter statistical mechanics, for
$\alpha=1-q$ and $\beta=0$ whilst, for instance, for
$\alpha=\beta=\kappa$, we obtain a new one-parameter subclass of
deformed logarithms (Kaniadakis 2005).
Therefore, the question of the analytical derivation
of the deformed parameter arise also in other context than the $q$-statistics.
To cite an example, the deformed parameter $\kappa$ has been determined implicitly in
Kaniadakis (2002, 2005) by considering the black-hole entropy, obtaining the relation
\begin{equation}
\ln_{_{\{k\}}}(M)={2\over3} \ ,
\end{equation}
where $M$ is the mass of the black-hole. A similar relation has
also been derived  in the case of the $q$-statistics.


\sect{Application to stellar plasma and equilibrium time}

Because many nuclear reactions in the stellar burning
core proceed by way of quantum penetration of a
high Coulomb barrier, their cross sections grow exponentially
with energy. Therefore, thermal averages do
not probe the average energy of the distribution ($kT$),
but its high-energy tail. Consequently, rates are very sensitive
to a relatively small part of the distribution.

The reasons of such deviations from the exponential tail have been
discussed elsewhere (Coraddu et al. 1999; Ferro and Quarati 2005) and include
the presence of distributions of random microfields
larger than a critical value and the quantum uncertainty between energy
and momentum of effective particles which leads to a non-exponential
momentum distribution even in presence of an exponential energy
distribution. Analytical expressions for these effects have already
been presented in this paper.

In addition, screening affects fusion rate in stars not only by
lowering the Coulomb barrier and, therefore, increasing the penetration,
but also by modifying the distribution. In fact, screening induces
effective nonlocal and retarded interactions between colliding particles,
which can not be treated as pure Coulomb collisions.
In this respect, since the number of particles inside the Debye-H\"{u}ckel
sphere is too small one should also introduce modifications of the
usual Debye-H\"{u}ckel screening.

Finally, there could be further non-Gaussian effects due to the
fact that distributions can be out of equilibrium. Indeed,
nuclear burning in stellar cores is a non-equilibrium
process with the corresponding ion distributions that, in principle,
depart from exponential. However, the deviation from the pure
exponential due to the reacting ions is completely negligible for
stellar structure. As Bahcall et al. (2002) have
argued, this effect is proportional to the ratio
of the collision and nuclear reaction times
\begin{equation}
\delta={\tau_{\rm Coul}\over\tau_{\rm nucl}} \ ,
\end{equation}
which is usually very small; for instance, this ratio can
be estimated of the order of $\delta\simeq10^{-(20\div28)}$
for protons in the sun. Minor and faster reactions can yield
larger ratios.

This line of arguing is based on the assumption that only two
timescales are relevant for the distribution. In fact, collective
modes on several scales are present in nuclear plasmas that have
different timescales. For instance, the ratio
$\nu_1/\nu_{\rm Coul}$, where $\nu_1$ is the collision
frequency from the so-called reinforced Coulomb scattering, is not
at all negligible, possibly leading to non-Gaussian behavior.

Therefore, one should carefully assess what is the equilibrium
relevant to the specific problem at hand (nuclear fusion,
radiative recombination, diffusion, etc.) and
also consider, as also argued in Lavagno and Quarati (2000, 2006), that
non-Gaussian behavior in plasmas arises from all collective behaviors
discussed above not only from non-equilibrium.

We remind that astrophysical and heavy-ion plasmas require relativistic
extensions of the $q$-formalism (Lavagno 2002).
The reader might be also interested to some recent applications of
nonextensive statistical mechanics, where a physical meaning of $q$
is introduced: to astrophysics by Jiulin and Yeli (2008a,2008b)
to heavy ions by Wilk and collaborators (Osada and Wilk 2007).

\sect{Conclusion}
Non-Gaussian distribution functions appear in interacting particle
systems with nonlinear, nonlocal, metastable, or long-memory systems;
plasmas are typical and important example of physical context
where such distributions appear.

Modified statistical mechanics,
such as the one introduced by Tsallis, give consistent frameworks
that naturally yield distributions that deviate from the Maxwellian.
For many purposes a single effective parameter, such as the
entropic parameter $q$, is sufficient to characterize the
non-Gaussian behavior of the system. This parameter can
always, in principle, be derived from the underlying microscopical
dynamics, even if complete dynamic
calculations are very difficult for most of the systems.

Even when the calculation from first principle is possible, the
answer does not need to be a simple analytical formula, but one
often can find numerically the parameter(s) that describe(s) the
distribution; nevertheless the result is predictive, since the
same number applies to many distributions for similar systems.

In this paper, we have reported a few cases where there exist
an analytical formula for the entropic parameter $q$ in terms of
properties of the systems. All these cases involve particles that
are properly correlated. Other approaches are certainly possible, but
nonextensive $q$ statistical mechanics appears particularly well-suited
to describe such complex systems, particularly astrophysical and space systems.

At last, we recall that $q$-Gaussian distributions, in addition to
provide a very powerful effective description of complex systems,
have also a predictive power. For instance, Lutz (2003) proposed
an experiment on optical lattices; such an experiment was later
performed (Douglas et al. 2006) confirming the expectation. We have also
proposed (Maero et al. 2006) an experiment to look for a cut-off in the
energy distribution of gammas in electron recombination that could
confirm deviations and possibly measure the relevant parameter of the
distribution. To obtain predictive results in the astrophysics and space
science field is the challenge of the next future.

\end{document}